\documentclass[sigconf]{acmart}
\AtBeginDocument{%
  }

\setcopyright{acmlicensed}
\copyrightyear{2018}
\acmYear{2018}
\acmDOI{XXXXXXX.XXXXXXX}
\acmConference[Conference acronym 'XX]{Make sure to enter the correct
  conference title from your rights confirmation email}{June 03--05,
  2018}{Woodstock, NY}
\acmISBN{978-1-4503-XXXX-X/2018/06}

\usepackage{graphicx}
\usepackage{booktabs} 
\usepackage{amsmath}  

\usepackage{array}    
\usepackage{longtable}
\usepackage{tabularx} 
\usepackage{hyperref}
\usepackage{svg}
\usepackage{float}
\newcolumntype{R}{>{\raggedleft\arraybackslash}X}
\usepackage{ragged2e}   
\usepackage{siunitx}
\usepackage{cleveref}
\usepackage{hyperref}
\usepackage{url}
\usepackage{enumitem}

\begin{document}
\title{CA-DEL: An Open Multi-Target, Multi-Modal Benchmark for Learning from DNA-Encoded Library Screens}

\author{Mutian He}
\authornote{Both authors contributed equally to this work.}
\affiliation{%
  \institution{Macao Polytechnic University}
  \state{Macao}
  \country{China}
}

\author{Hanqun Cao}
\authornotemark[1]
\affiliation{%
  \institution{The Chinese University of Hong Kong}
  \state{Hong Kong}
  \country{China}
}

\author{Cheng Tan}
\affiliation{%
  \institution{Shanghai Artificial Intelligence Laboratory} 
  \state{Shang Hai}
  \country{China}     
}

\author{Zijun Gao}
\affiliation{%
  \institution{The Chinese University of Hong Kong}
  \state{Hong Kong}
  \country{China}
}

\author{Xiaojun Yao}
\affiliation{%
  \institution{Macao Polytechnic University}
  \state{Macao}
  \country{China}
}

\author{Chunbin Gu}
\authornote{Corresponding author.}
\affiliation{%
  \institution{The Chinese University of Hong Kong}
  \state{Hong Kong}
  \country{China}
}

\author{Pheng-Ann Heng}
\affiliation{%
  \institution{The Chinese University of Hong Kong}
  \state{Hong Kong}
  \country{China}
}
\renewcommand{\shortauthors}{He et al.}



\received{20 February 2007}
\received[revised]{12 March 2009}
\received[accepted]{5 June 2009}

\begin{abstract}
The success of machine learning in drug discovery hinges on learning the relationship between a chemical structure and its biological activity. While DNA-Encoded Library (DEL) technology can generate the massive datasets required for this task, its primary signal---sequencing read counts---is an indirect and often noisy proxy for true molecular binding affinity. To address the scarcity of public benchmarks for developing robust models that can overcome this data challenge, we introduce CA-DEL, a multi-dimensional public benchmark featuring screens against three homologous carbonic anhydrase isoforms. While recent benchmarks like KinDEL have introduced 3D poses for kinase targets, CA-DEL distinguishes itself by focusing on the selectivity challenge among homologous Carbonic Anhydrase isoforms (CAII, CAIX, CAXII). Unlike benchmarks relying solely on noisy enrichment scores, CA-DEL integrates a rigorous validation set of experimentally determined binding affinities ($K_i$) from ChEMBL, establishing a critical Sim-to-Real evaluation paradigm: training on noisy DEL screens and testing on high-fidelity biophysical data.

\end{abstract}

\maketitle

\section{Introduction}
\begin{figure}[t]
    \centering
    \includegraphics[width=0.9\linewidth]{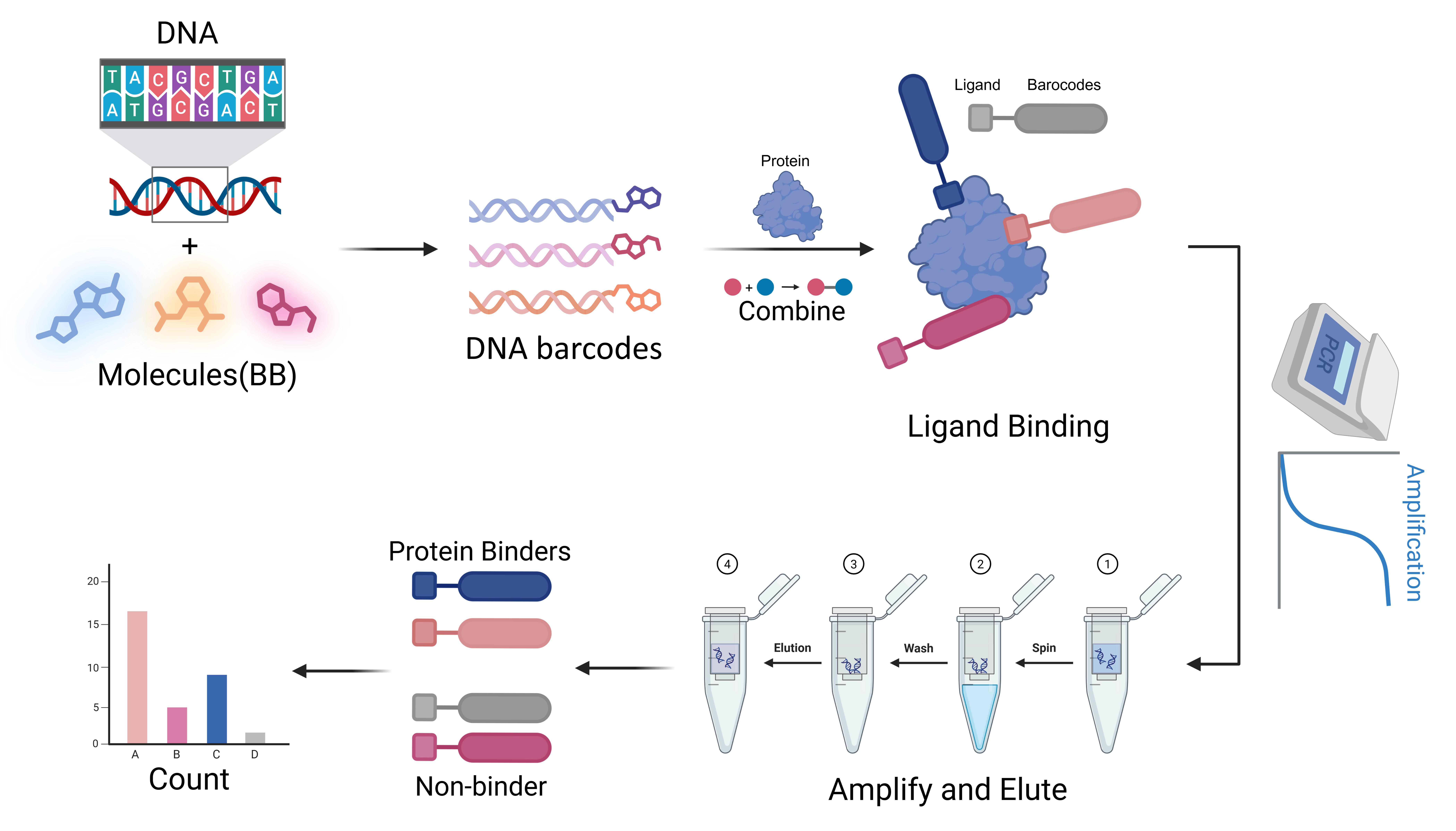}
    \caption{Schematic diagram of the DNA-encoded library (DEL) screening process.}
    \label{fig:del_workflow} 
\end{figure}

DNA-Encoded Library (DEL) technology has become a cornerstone of modern drug discovery, enabling unprecedented screening throughput of libraries containing billions of unique molecules covalently linked to DNA barcodes(Figure~\ref{fig:del_workflow}), \citep{needels1993generation, brenner1992encoded}. However, the primary experimental output—high-throughput sequencing read counts—is not a direct measure of binding affinity but rather a noisy proxy confounded by non-specific binding, synthesis impurities, and PCR amplification preferences~\cite{dumelin2006selection,wichert2024challenges}. Effective DEL analysis therefore requires computational models capable of denoising, debiasing, and ranking molecules from these weak, indirect signals~\cite{ma2021regression, lim2022machine,gu2024unlocking,cao2024ranking,iqbal2025evaluation}.

While recent advancements have established strong baselines, several avenues remain for further optimization in molecular learning. First, molecular recognition is inherently three-dimensional. Although 2D representations are computationally efficient, incorporating 3D geometric and stereochemical contexts can further enhance predictive fidelity, particularly when subtle structural variations drive binding outcomes~\cite{shmilovich2023dock}. Second, existing benchmarks have made significant contributions to the field: KinDEL provides an extensive resource for kinase hit finding~\cite{chen2024kindel}, while BELKA offers an unprecedented scale of 133M molecules, demonstrating the power of large-scale screening~\cite{quigley2024belka}. 

Complementing these foundational large-scale datasets, our work focuses on Carbonic Anhydrase (CA) to offer a specialized testbed. Instead of relying solely on binary labels, this approach bridges the gap between high-throughput screening and lead optimization by effectively integrating DEL enrichment signals with experimentally measured binding affinities. This granularity is essential for modeling the complex relationship between screening readouts and physical binding properties. Third, the clinical imperative extends beyond identifying binders to ensuring high target selectivity. This is particularly critical for highly homologous protein families like CA, where discerning mechanisms are required to navigate the complex landscape of selective inhibitor design.

To address these limitations, we introduce CA-DEL, a multi-dimensional public benchmark dataset designed specifically for DEL data analysis that advances beyond existing resources in scope, modality, and biological relevance. 
Our contributions include: 
\begin{itemize}[leftmargin=*]
    \item \textbf{Multi-target selectivity benchmark.} We evaluate selectivity prediction against highly homologous carbonic anhydrase isoforms (CAII, CAIX, CAXII), representing a clinically relevant challenge where inhibitor selectivity must be achieved between isoforms with conserved catalytic mechanisms.
    \item \textbf{Multi-modal molecular representations.} We integrate traditional 2D molecular topology with systematically generated 3D protein-ligand conformations for over 200K compounds, enabling a large-scale evaluation of geometric deep learning approaches on CA-DEL data.
    \item \textbf{Ground-truth validated OOD challenge.} We design an Out-of-Distribution task that tests generalization from noisy DEL screening data (enrichment factors) to precise ChEMBL binding affinities ($K_i/K_d$ values), spanning distinct chemical spaces that mirror real-world hit-to-lead optimization.
    \item \textbf{Practical evaluation metrics.} We introduce Top-N hit rate analysis that directly assesses model utility in resource-constrained discovery campaigns, moving beyond traditional correlation metrics to measure practical discovery value.
\end{itemize}
Unlike existing benchmarks that focus primarily on binary classification or single-target activity prediction, CA-DEL provides a comprehensive platform for developing and evaluating 3D-aware, selectivity-focused machine learning models on DEL data, potentially catalyzing advances in geometric deep learning approaches for structure-based drug design.

This benchmark design addresses the clinical reality where models trained on initial screening data must generalize to lead-optimized compounds occupying different chemical space, while forcing models to learn isoform-specific binding features rather than generalized motifs. The Top-N evaluation paradigm measures what matters most to discovery scientists: the percentage of true hits among top-ranked compounds.

\begin{figure}[t]
    \centering
    \includegraphics[width=0.9\linewidth]{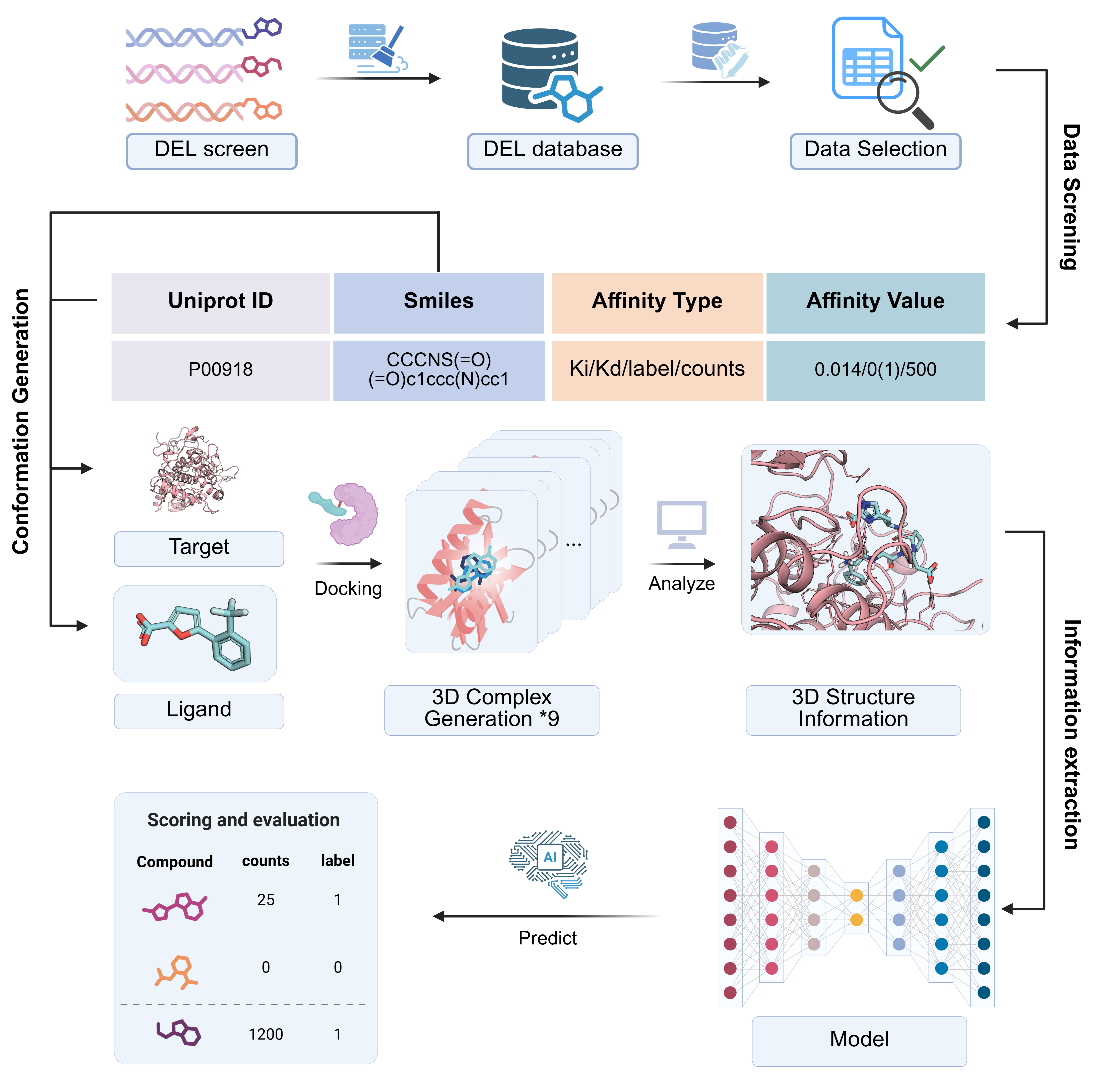}
    \caption{\textbf{Schematic overview of the proposed structure-based deep learning framework for DEL analysis.} 
    The workflow is organized into three distinct phases: 
    (1) \textbf{Data Screening}, where raw DEL screening data is processed to align target Uniprot IDs with ligand SMILES and affinity metrics; 
    (2) \textbf{Conformation Generation}, which utilizes molecular docking to generate 3D protein-ligand complexes. An intermediate \textit{analysis} step is employed to evaluate the docking results and verify the correctness of the binding conformations before finalizing the 3D structure information; 
    and (3) \textbf{Information Extraction}, where the validated structural features are fed into a neural network model to predict bioactivity scores and evaluate compound labels.}
    \label{fig:workflow}
\end{figure}

\section{Related Work}
\subsection{DEL Analysis}
The successful application of DEL technology has generated vast quantities of high-dimensional, high-noise screening data, thereby driving the continuous evolution of computational methods. 

Early approaches focused on hit identification and initial structure-activity relationship (SAR) analysis, relying on methods including physics-based molecular docking simulations \citep{wang2015accurate}, quantitative structure-activity relationship (QSAR) models, and metrics specific to DEL enrichment such as data aggregation \citep{satz2016simulated},

Enrichment Factor calculation, and standardized z-scores \citep{faver2019quantitative}. While intuitive, these methods possess limited capabilities in handling complex, non-linear structure-activity relationships. To overcome these limitations, machine learning models such as gradient boosting machines, random forests, and support vector machines \citep{mccloskey2020machine, li2007machine, ballester2010machine} elevated DEL data into a powerful platform for driving predictive modeling. More recently, the rise of deep learning, particularly the application of Graph Neural Networks (GNNs) \citep{ma2021regression}, has further enhanced the modeling of complex molecular structures.

Nevertheless, the aforementioned methods predominantly rely on 2D topological information. This represents a fundamental limitation, as the physical essence of molecular recognition occurs in three-dimensional space. Consequently, the effective integration of 3D structural information into predictive models has emerged as a key research frontier. Recent innovations include DEL-Dock, a multi-modal neural network that combines 3D conformational information with 2D topology \citep{shmilovich2023dock}; sparse learning methods that address noise originating from truncated products and sequencing errors \citep{komar2020denoising}; and DEL-ranking \citep{cao2024ranking}, which leverages multiple 3D conformations through a designed ranking-based loss function to more effectively correct read count distributions while concurrently addressing issues of distributional noise and shift.

\subsection{Existing Benchmarks for DEL Analysis}
However, the development of advanced algorithms has long been constrained by a core bottleneck: the scarcity of public benchmark datasets. This situation severely restricts the development and fair comparison of novel computational methods. Early public datasets were modest in scale, such as those for casein kinase and an early dataset for CA \citep{ballester2010machine,iqbal2025evaluation}. They were instrumental in validating specific analytical methods like probabilistic loss functions, demonstrating the significant value of specialized datasets in catalyzing targeted methodological advancements. Subsequently, the advent of large-scale benchmarks propelled the field to new heights. KinDEL\citep{chen2024kindel}, one of the first large-scale public datasets, provided data for over 80 million compounds against two kinase targets (MAPK14 and DDR1), including both on-DNA enrichment data and off-DNA biophysical validation data, offering a valuable resource for studying the concordance between DEL screening and traditional assays. More recently, the BELKA dataset, released as part of a Kaggle competition, comprised approximately 133 million molecules against three targets (BRD4, sEH, and HSA). By formulating the problem as a binary classification task (binding vs. non-binding), it significantly lowered the barrier to entry for the broader machine learning community, catalyzing the application of a diverse range of classification algorithms to the DEL hit-finding problem.
Current benchmarks for DEL analysis generally lack fine-grained 3D structural information and unified benchmark tasks. This limitation makes it difficult to train and compare models that aim to leverage spatial information for improved predictive accuracy.

\section{Dataset Construction and Description}

This section details the construction methodology of the CA-DEL dataset (Figure~\ref{fig:workflow}), designed to establish a benchmark for evaluating machine learning models on selective protein-ligand binding prediction. We describe target protein selection, small molecule data curation from heterogeneous sources, multi-modal molecular representation generation protocols, and statistical properties of the final dataset.

\begin{table}[t]
    \centering
    \small 
    
    \caption{\textbf{Overview of the CA-DEL dataset composition.} 
    Adapted to single-column layout.
    Top: \textbf{Training sets} derived from DEL selections (Enrichment Data).
    Bottom: \textbf{Test sets} consisting of bioactivity data ($K_i$) from ChEMBL.
    Note: "Split" and "Purpose" columns are implied by the grouping.}
    \label{tab:dataset_composition}

    \resizebox{\linewidth}{!}{
        \begin{tabular}{@{}l l l l c l@{}}
            \toprule
            \multicolumn{6}{c}{\textbf{Training Sets (Source: DEL Selections)}} \\
            \midrule
            \textbf{Target} & \textbf{Uniprot} & \textbf{PDB ID} & \textbf{Source} & \textbf{Compounds} & \textbf{Data Type} \\
            \midrule
            CAII  & P00918 & 3p3h, 5doh & CAS-DEL   & 127,500 & Enrichment \\
            CAIX  & Q16790 & 2hkf, 5fl4 & DOS-DEL-1 & 108,528 & Enrichment \\
            CAXII & O43570 & 4kp5, 4ht2 & CAS-DEL   & 127,500 & Enrichment \\
            \bottomrule
        \end{tabular}
    }

    \vspace{0.2cm} 

    \resizebox{\linewidth}{!}{
        \begin{tabular}{@{}l l l l c l@{}}
            \toprule
            \multicolumn{6}{c}{\textbf{Test Sets (Source: ChEMBL)}} \\
            \midrule
            \textbf{Target} & \textbf{Uniprot} & \textbf{PDB ID} & \textbf{Source} & \textbf{Compounds} & \textbf{Data Type} \\
            \midrule
            CAII  & P00918 & 3p3h, 5doh & ChEMBL & 6,396 & $K_i$ \\
            CAIX  & Q16790 & 2hkf, 5fl4 & ChEMBL & 3,323 & $K_i$ \\
            CAXII & O43570 & 4kp5, 4ht2 & ChEMBL & 2,689 & $K_i$ \\
            \bottomrule
        \end{tabular}
    }
\end{table}

\subsection{Target Protein and Small Molecule Selection}
CA-DEL evaluates model generalization across two critical dimensions: \textit{biological selectivity} between homologous targets and \textit{domain generalization} across data distributions. We selected three human carbonic anhydrase isoforms: CAII (ubiquitous anti-target), CAIX and CAXII (cancer-specific targets). Despite high active-site homology, these proteins differ significantly in physiological roles, creating a challenging multi-task learning objective where models must learn fine-grained structural features governing isoform selectivity.

The dataset employs heterogeneous data sources to probe model robustness. Training data originates from two DELs with distinct chemical spaces: CAS-DEL library (127,500 compounds) ~\citep{hou2023machine} for CAII/CAXII and DOS-DEL-1 library (108,528 compounds) ~\citep{gerry2019dna} for CAIX (Table~\ref{tab:dataset_composition}). Validation and test sets source entirely from ChEMBL ~\citep{gaulton2012chembl}, comprising drug-like molecules (CAII: 6,396; CAIX: 3,323; CAXII: 2,689) with precise $K_i$ measurements. This deliberate distributional shift—spanning both chemical space and activity label modality—constitutes a realistic OOD generalization challenge that emulates real-world transfer from high-throughput screening to lead optimization. Data field descriptions are provided in Tables~\ref{tab:appendix_data_fields_cas}, \ref{tab:appendix_data_fields_dos}, and~\ref{tab:appendix_data_fields_chembl}.

\subsection{Generation of Multi-Modal Molecular Representations}

Beyond high-throughput sequencing read counts from DEL wet-lab experiments, we established a systematic pipeline for generating multi-modal representations rich in 3D structural information.

For each target protein, two high-resolution crystal structures were selected from the Protein Data Bank (PDB) (CAII: \texttt{3p3h}, \texttt{5doh}; CAIX: \texttt{2hkf}, \texttt{5fl4}; CAXII: \texttt{4kp5}, \texttt{4ht2}) and subjected to standard preparation protocols using PDB2PQR and PROPKA to account for protein conformational flexibility. For each small molecule, an initial 3D conformation was generated from its SMILES string using RDKit, followed by energy minimization with the MMFF94 force field.

Since docking scoring functions often fail to identify the true binding mode as the top-ranked result, we used SMINA \citep{koes2013lessons} to generate ensembles of up to nine plausible binding poses for each ligand-protein pair. This approach increases the probability of capturing the correct binding mode, which typically resides among top-scoring poses. Docking was constrained to a 22.5 Å cubic search space defined by known co-crystallized ligand positions. By providing pose ensembles, our pipeline becomes more resilient to upstream scoring errors and furnishes more physically complete inputs ~\citep{shamsian2023benchmarking,wang2003comparative,ferrara2004assessing}. This strategy replaces single, potentially incorrect binding mode estimates with robust discrete approximations of the conformational posterior distribution, enabling models to learn more nuanced and generalizable protein-ligand interaction representations.

\subsection{Dataset Statistics and Analysis}

The CA-DEL dataset was constructed as a rigorous benchmark to address two fundamental challenges in computational drug discovery: learning from noisy ~\citep{kuai2018randomness,montoya2025widespread,li2007machine,lim2022machine,komar2020denoising}, high-throughput screening data and generalizing to OOD, drug-like chemical matter ~\citep{liu2021towards,tossou2024real,shi2025out}) The following sections detail the dataset's strategic design and insights from preliminary analysis.

\begin{figure}[t]
    \centering
    \includegraphics[width=0.9\linewidth]{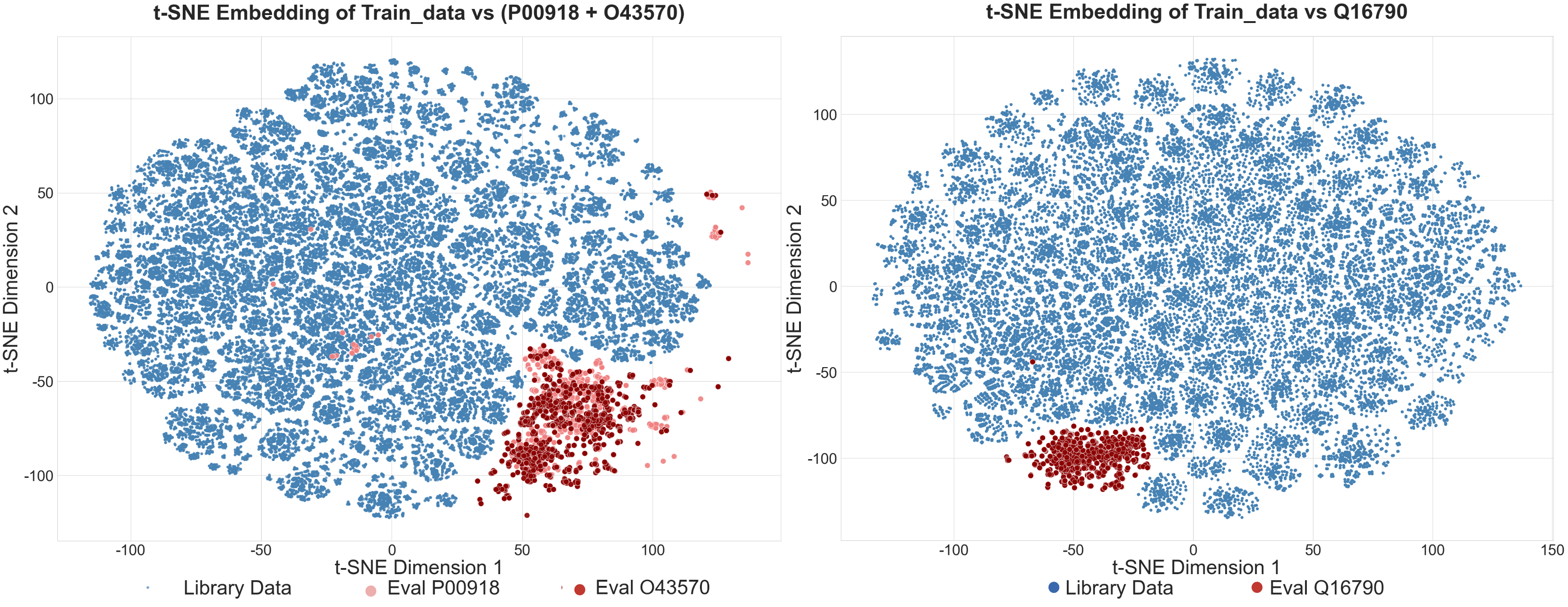}
    \caption{t-SNE visualization of the chemical space. The clear separation between the DEL training set (blue point cloud) and the ChEMBL validation/test set (red star-shaped cluster) highlights the significant distributional shift engineered into the benchmark.}
    \label{fig:tsne}
\end{figure}

\textbf{Significant distributional shift between training and test sets.} 
We deliberately introduced a significant distributional shift between the training and test sets to simulate the real-world progression from screening hits to optimized lead compounds. This multifaceted domain gap covers disparities in chemical structure and physicochemical properties. A t-SNE projection (Figure~\ref{fig:tsne}) starkly visualizes this structural chasm, showing that the training data (combinatorial libraries) and evaluation sets (ChEMBL) occupy distinct, non-overlapping regions of chemical space. Consequently, the model must learn to extrapolate generalizable biophysical principles rather than interpolate library-specific structural motifs.

\begin{figure}[t]
    \centering
    \includegraphics[width=0.92\linewidth]{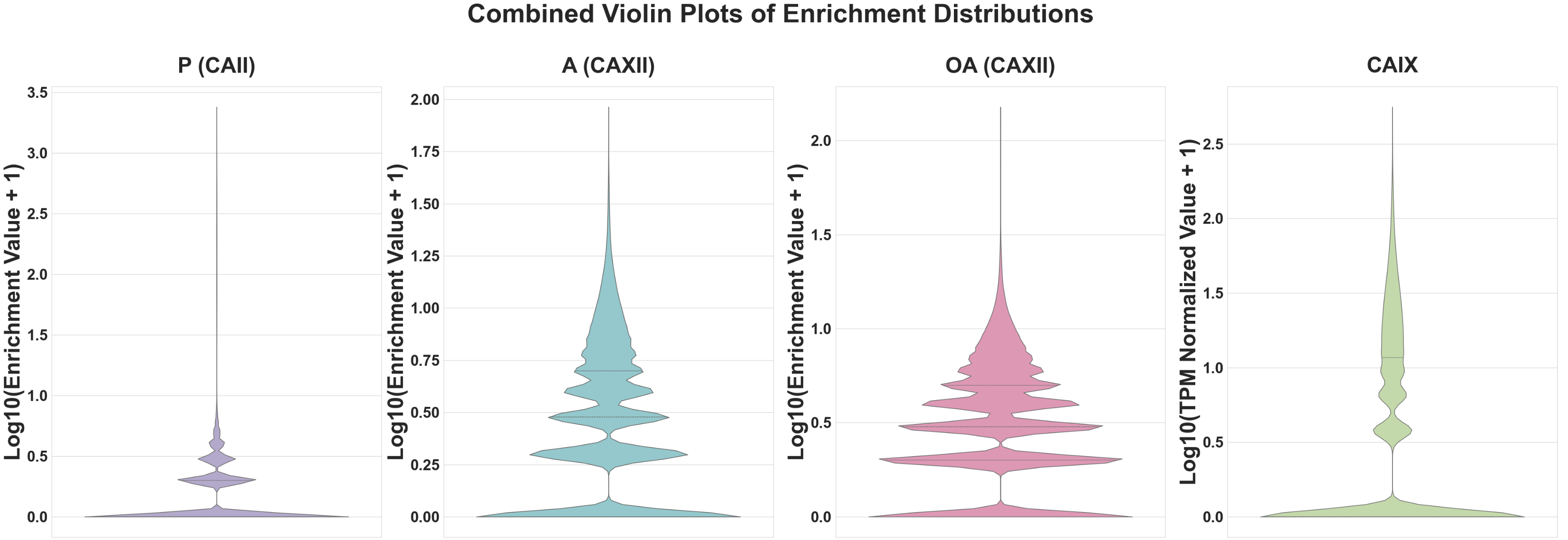}
    \caption{Distribution of enrichment values from the DEL screening data (training set). The violin plot illustrates a typical long-tail distribution, where most library compounds are inactive and a small fraction constitutes potential hits.}
    \label{fig:vio}
\end{figure}

\begin{figure*}[t]
    \centering
    \includegraphics[width=0.85\textwidth]{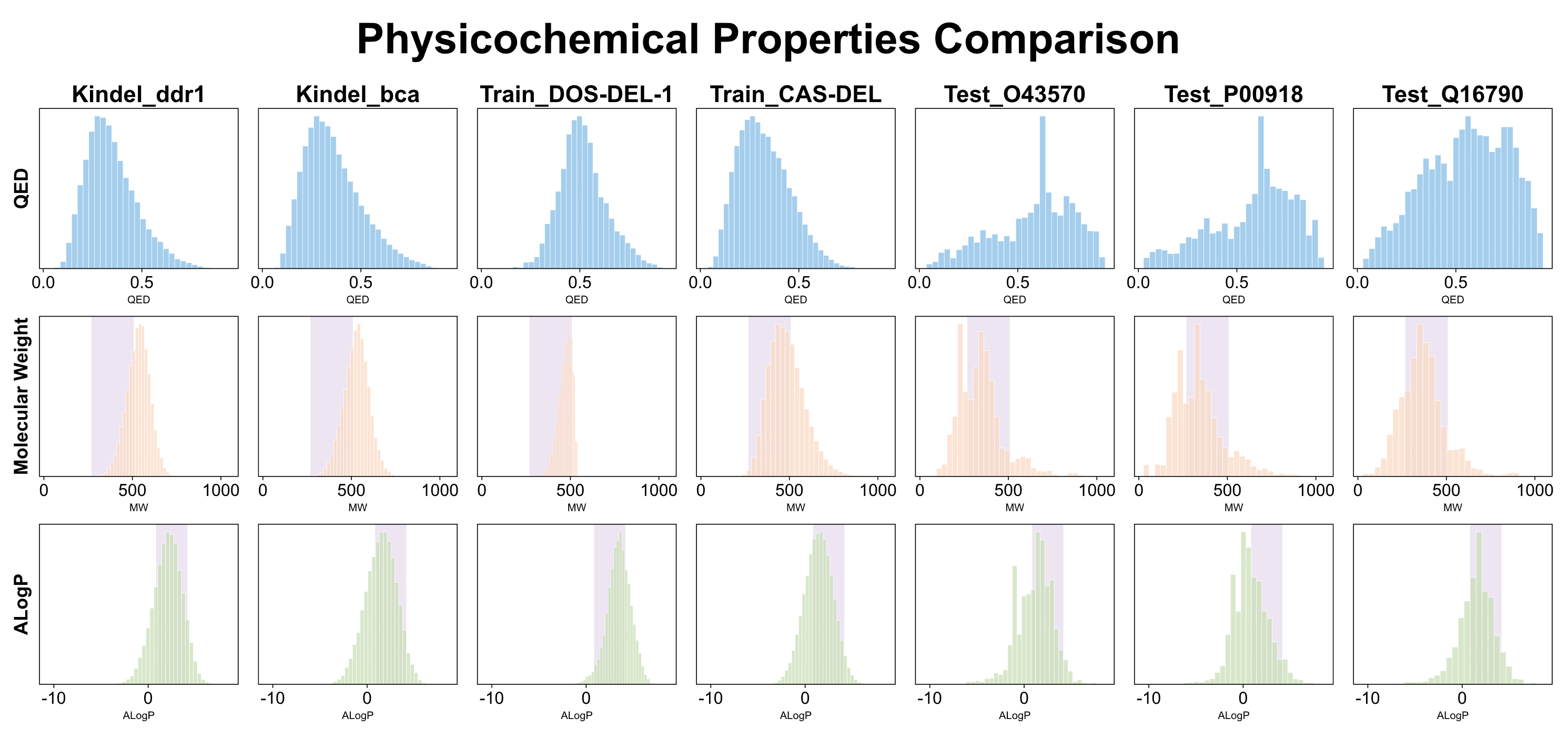}
    \Description{Histograms comparing QED, LogP, and Molecular Weight across DOS-DEL-1, CAS-DEL, KinDEL, and test sets.}
    
    \caption{
        Comparison of key physicochemical properties (QED, LogP, and Molecular Weight) across distinct ligand datasets, which includes source datasets (DOS-DEL-1 and CAS-DEL), benchmark dataset KinDEL~\citep{chen2024kindel}, and our curated test sets. The light blue areas mark the 10th and 90th percentiles computed for all the FDA approved oral new chemical entities, as reported by ~\citep{shultz2018two}. QED: quantitative estimate of druglikeness~\citep{bickerton2012quantifying}.
    }
    \label{fig:physchem_props1}
\end{figure*}

This structural divergence is mirrored by a systematic shift in physicochemical properties (Figures~\ref{fig:physchem_props1} and~\ref{fig:physchem_props2}). Training molecules resemble initial hits, with lower Quantitative Estimate of Drug-likeness (QED) and higher molecular weights. In contrast, the test sets are more drug-like, with higher QED values and optimized weights. This complex, multi-dimensional gap across various descriptors presents a rigorous test, challenging the model to avoid learning spurious, library-specific correlations.

\begin{figure}[t]
    \centering
    \includegraphics[width=0.45\textwidth]{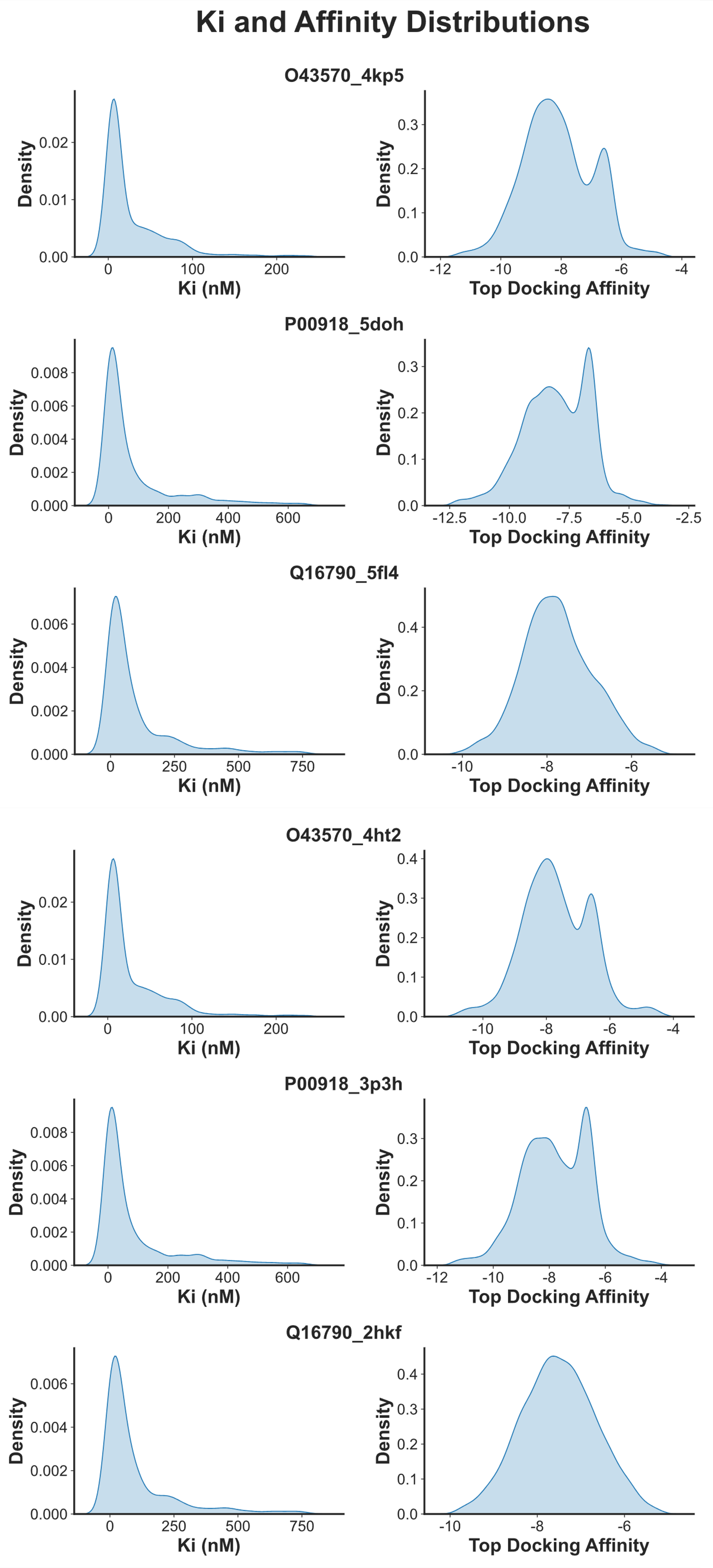} 
    \Description{Distribution histograms showing binding affinity scores.}
    \caption{Binding affinity distribution and docking scores for ChEMBL test compounds.}
    \label{fig:docking_output}
\end{figure}

The model is trained on noisy, long-tailed enrichment factors derived from competitive DEL selection experiments(Figure~\ref{fig:vio}), which are inherently a relative and semi-quantitative measure of binding preference. In stark contrast, the evaluation is performed against precise, absolute biophysical measurements such as binding affinities ($K_i/K_d$) or computationally derived docking scores, whose distributions are shown in Figure~\ref{fig:docking_output}.

This necessitates that the model not only navigates the covariate shift in molecular features but also performs a complex translation from a noisy, relative experimental signal to a quantitative, absolute biophysical value. Success under these conditions would strongly imply that the model has learned a robust and transferable latent representation that captures the fundamental physics of protein-ligand interactions, effectively distilling the true binding signal from the experimental noise and artifacts inherent in the training data.

\subsubsection{The Biological Selectivity Challenge}

\textbf{Predicting selectivity across highly homologous CA isoforms.} The benchmark targets CA isoforms II, IX, and XII, where CAII represents a ubiquitous anti-target and CAIX/CAXII are validated cancer targets. These proteins exhibit high structural similarity in their active sites, making selective inhibitor design challenging~\citep{mboge2018selective,genis2009design,alterio2009crystal}. The primary challenge stems from the conserved catalytic zinc ion and subtle structural variations, often single amino acid substitutions, that dictate isoform-specific binding. This framework establishes the task as multi-target representation learning, where effective models must identify minute structural and chemical distinctions determining selectivity. The benchmark directly evaluates a model's capacity to guide lead optimization, where achieving selectivity is essential for therapeutic success.

\subsubsection{Generation and Validation of Physically Realistic Binding Poses}
\textbf{A cornerstone of our methodology is the generation of physically realistic binding poses whose structural accuracy is rigorously validated against experimental data.} As demonstrated by our re-docking analysis, the generated poses successfully recapitulate the critical binding interactions observed in experimental co-crystal structures (Figure~\ref{fig:re-docking}). By representing each ligand-protein complex with an ensemble of these validated, low-energy conformations, we decouple model performance from the known unreliability of docking scoring functions ~\citep{mukherjee2010docking,scardino2021combination,ferrara2004assessing}. This strategy forces the model to learn "consensus" interaction features that are robust to minor conformational shifts, encouraging the development of representations that better reflect physical reality and exhibit superior generalization.

\section{Results and Analysis}
To establish a comprehensive performance baseline on the CA-DEL benchmark, we evaluated a range of models, from simple physicochemical descriptors to advanced, multi-modal deep learning architectures. The detailed results are presented in Table~\ref{tab:main_results_part1} and Table~\ref{tab:main_results_part2}. Our analysis is structured across multiple dimensions—examining the performance from the perspectives of model methodology, dataset characteristics, and the evaluation metrics themselves—to provide a multi-faceted understanding of the benchmark's challenges and the key drivers of model success.

The primary evaluation task assesses a model's ability to rank compounds by binding potential through correlation with experimental read counts. Performance is measured using Spearman's rank correlation coefficient ($\rho$), defined as:

\begin{equation}
\rho = 1 - \frac{6\sum_{i=1}^{n} d_i^2}{n(n^2-1)}
\end{equation}

where $d_i$ represents the difference between the ranks of corresponding observations, and $n$ is the number of observations. Based on rank information, we provided Top-N hit rate comparisons to simulate practical utility for scientific discovery and conducted zero-shot generalization experiments for more stringent conditions.

\subsection{Analysis by Dataset Characteristics and Evaluation Metrics}

The CA-DEL dataset was intentionally designed to probe fundamental challenges in drug discovery, and the results reflect how different models cope with these hurdles. A primary driver of the observed performance differences is the pronounced Out-of-Distribution (OOD) distributional shift between the DEL-sourced training set and the ChEMBL-sourced test set. Models that merely memorize statistical artifacts of the training library are destined to fail on this task. The poor performance of simple physicochemical baselines and traditional machine learning models on the SubSp metric, which measures correlation with true binding affinity, demonstrates their failure to generalize across this domain shift. In contrast, the success of the 3D deep learning models, such as DEL-Dock and DEL-Ranking, suggests they learn more fundamental and transferable representations of biophysical interactions rather than library-specific patterns. The performance hierarchy among different algorithm types aligns with conventional understanding in computational drug discovery, serving as a crucial validation of the proposed CA-DEL dataset's ability to meaningfully differentiate algorithmic capabilities and capture the inherent complexity of real-world drug discovery scenarios.

Our dual-metric evaluation framework serves as a powerful diagnostic tool for this purpose. Performance is assessed using two Spearman's rank correlation coefficients: Sp, which measures the correlation with the noisy DEL read counts, and SubSp, which measures the correlation with the ground-truth $K_i$ values on the ChEMBL subset. A model that simply overfits to the training data might achieve a high Sp but will fail to generalize, resulting in a low SubSp. Conversely, a model that successfully denoises the training signal will learn the underlying biophysical relationships, resulting in a strong SubSp even if its Sp is not perfect.

\textbf{A critical aspect of our evaluation is the interpretation of the SubSp metric.} Since the model is trained to output a higher score for a more promising compound (e.g., from higher enrichment values), while the ground-truth labels are binding affinities like $K_i$ where a \textit{lower} value indicates stronger binding, a strong \textit{negative} correlation (i.e., $\rho \to -1$) signifies superior model performance. This indicates that the model correctly ranks compounds with stronger binding affinity higher.

\subsection{Analysis by Model Complexity and Modality}
The results reveal a clear performance hierarchy that directly correlates with model sophistication and the nature of the input data in Table~\ref{tab:main_results_part1} and Table~\ref{tab:main_results_part2}. This stratification underscores the complexity of the task and highlights the limitations of traditional approaches.

\textbf{Physicochemical and Heuristic Baselines.} Simple models utilizing single physicochemical properties, such as Molecular Weight or benzene ring presence, display poor and inconsistent performance across all targets. Their Spearman's rank correlation coefficients for both read counts (Sp) and true binding affinities (SubSp) approach zero, demonstrating virtually no meaningful predictive capacity for either DEL enrichment signals or genuine biological activity. This demonstrates that simplistic heuristics cannot address the intricate structure-activity relationships embedded within the dataset.

\textbf{Classical Docking and Traditional Machine Learning.} The performance of classical molecular docking (Vina Docking) is also limited, serving as a crucial reference point. For instance, on the '5doh' target, its SubSp is merely $-0.017 \pm 0.003$, showing almost no correlation with the true affinities. This result validates our dataset's premise that relying solely on classical docking scores is insufficient and motivates the development of more sophisticated, data-driven models.  While traditional machine learning models like Random Forest offer some improvement, they lack the consistency and robustness required for this challenging OOD task.

\textbf{Advanced Deep Learning: The Superiority of 3D Representations.} The results unequivocally demonstrate the superiority of deep learning-based approaches, particularly those that integrate 3D structural information. Multi-modal models like DEL-Dock and DEL-Ranking consistently achieve the strongest performance, especially on the critical SubSp metric, which measures correlation with true binding affinity.For the CAIX targets, these models achieve SubSp values as strong as $-0.308$ and $-0.323$ for '2hkf' and '5fl4' respectively. This performance significantly surpasses that of 2D-based models and all other baselines, highlighting the critical importance of leveraging 3D protein-ligand interaction information to successfully generalize from noisy DEL data to OOD chemical space of the ChEMBL test set.

\begin{table}[t] 
    \centering
    \small   
    \caption{Performance of Baseline Models (Part 1) adapted to single-column layout. 
    Targets are grouped into two sets.
    Top: Carbonic Anhydrase II (CAII) targets (5doh, 3p3h). 
    Bottom: Carbonic Anhydrase XII (CAXII) targets (4kp5\_A, 4kp5\_OA).
    Metric: Spearman's $\rho$ (Sp) and Subset Sp (SubSp).}
    \label{tab:main_results_part1}

    \resizebox{\linewidth}{!}{ 
        \begin{tabular}{@{}l c c c c@{}}
            \toprule
            & \multicolumn{2}{c}{\textbf{5doh}} & \multicolumn{2}{c}{\textbf{3p3h}} \\
            \cmidrule(lr){2-3} \cmidrule(lr){4-5} 
            \textbf{Metric} & Sp & SubSp & Sp & SubSp \\
            \midrule
            Mol Weight    & -0.250 & -0.125 & -0.250 & -0.125 \\
            Benzene       & 0.022  & 0.072  & 0.022  & 0.072  \\
            \midrule
            Vina Docking  & $-0.174$\tiny{$\pm 0.002$} & $-0.017$\tiny{$\pm 0.003$} & $-0.174$\tiny{$\pm 0.002$} & $-0.017$\tiny{$\pm 0.002$} \\
            RF-Enrichment & $0.108$\tiny{$\pm 0.006$}  & $0.100$\tiny{$\pm 0.015$}  & $-0.017$\tiny{$\pm 0.026$} & $-0.042$\tiny{$\pm 0.025$} \\
            RF-ZIP        & $-0.018$\tiny{$\pm 0.051$} & $-0.023$\tiny{$\pm 0.016$} & $0.027$\tiny{$\pm 0.139$}  & $-0.005$\tiny{$\pm 0.071$} \\
            Dos-DEL       & $-0.053$\tiny{$\pm 0.019$} & $-0.012$\tiny{$\pm 0.027$} & $-0.048$\tiny{$\pm 0.036$} & $-0.011$\tiny{$\pm 0.035$} \\
            DEL-QSVR      & $-0.175$\tiny{$\pm 0.021$} & $-0.092$\tiny{$\pm 0.033$} & $-0.228$\tiny{$\pm 0.021$} & $-0.171$\tiny{$\pm 0.033$} \\
            DEL-Dock      & $-0.181$\tiny{$\pm 0.075$} & $-0.085$\tiny{$\pm 0.061$} & $-0.255$\tiny{$\pm 0.009$} & $-0.137$\tiny{$\pm 0.012$} \\
            DEL-Ranking   & $-0.262$\tiny{$\pm 0.013$} & $-0.140$\tiny{$\pm 0.021$} & $-0.286$\tiny{$\pm 0.002$} & $-0.177$\tiny{$\pm 0.005$} \\
            \bottomrule
        \end{tabular}
    }

    \vspace{0.2cm} 

    \resizebox{\linewidth}{!}{ 
        \begin{tabular}{@{}l c c c c@{}}
            \toprule
            & \multicolumn{2}{c}{\textbf{4kp5\_A}} & \multicolumn{2}{c}{\textbf{4kp5\_OA}} \\
            \cmidrule(lr){2-3} \cmidrule(lr){4-5} 
            \textbf{Metric} & Sp & SubSp & Sp & SubSp \\
            \midrule
            Mol Weight    & -0.101 & 0.020 & -0.101 & 0.020 \\
            Benzene       & -0.054 & 0.035 & -0.054 & 0.035 \\
            \midrule
            Vina Docking  & $0.025$\tiny{$\pm 0.001$}  & $0.150$\tiny{$\pm 0.003$}  & $0.025$\tiny{$\pm 0.001$}  & $0.150$\tiny{$\pm 0.003$} \\
            RF-Enrichment & $-0.029$\tiny{$\pm 0.038$} & $-0.005$\tiny{$\pm 0.048$} & $-0.101$\tiny{$\pm 0.009$} & $-0.087$\tiny{$\pm 0.010$} \\
            RF-ZIP        & $0.035$\tiny{$\pm 0.094$}  & $-0.026$\tiny{$\pm 0.111$} & $0.006$\tiny{$\pm 0.095$}  & $-0.021$\tiny{$\pm 0.122$} \\
            Dos-DEL       & $-0.016$\tiny{$\pm 0.029$} & $-0.017$\tiny{$\pm 0.021$} & $-0.003$\tiny{$\pm 0.030$} & $-0.048$\tiny{$\pm 0.034$} \\
            DEL-QSVR      & $-0.004$\tiny{$\pm 0.178$} & $0.018$\tiny{$\pm 0.139$}  & $0.070$\tiny{$\pm 0.013$}  & $-0.076$\tiny{$\pm 0.116$} \\
            DEL-Dock      & $-0.242$\tiny{$\pm 0.011$} & $-0.263$\tiny{$\pm 0.012$} & $0.015$\tiny{$\pm 0.029$}  & $-0.105$\tiny{$\pm 0.034$} \\
            DEL-Ranking   & $-0.268$\tiny{$\pm 0.012$} & $-0.277$\tiny{$\pm 0.016$} & $-0.289$\tiny{$\pm 0.025$} & $-0.233$\tiny{$\pm 0.021$} \\
            \bottomrule
        \end{tabular}
    }
\end{table}

\begin{table}[t] 
    \centering
    \small 
    
    \caption{Performance of Baseline Models (Part 2) adapted to single-column layout. 
    Top: Carbonic Anhydrase XII (CAXII) targets (4ht2\_A, 4ht2\_OA). 
    Bottom: Carbonic Anhydrase IX (CAIX) targets (2hkf, 5fl4).
    Metric: Spearman's $\rho$ (Sp) and Subset Sp (SubSp).}
    \label{tab:main_results_part2}

    \resizebox{\linewidth}{!}{ 
        \begin{tabular}{@{}l c c c c@{}}
            \toprule
            & \multicolumn{2}{c}{\textbf{4ht2\_A}} & \multicolumn{2}{c}{\textbf{4ht2\_OA}} \\
            \cmidrule(lr){2-3} \cmidrule(lr){4-5} 
            \textbf{Metric} & Sp & SubSp & Sp & SubSp \\
            \midrule
            Mol Weight    & -0.101 & 0.020  & -0.101 & 0.020  \\
            Benzene       & -0.054 & 0.035  & -0.054 & 0.035  \\
            \midrule
            Vina Docking  & $-0.037$\tiny{$\pm 0.011$} & $0.092$\tiny{$\pm 0.011$}  & $-0.037$\tiny{$\pm 0.011$} & $0.092$\tiny{$\pm 0.011$} \\
            RF-Enrichment & $0.011$\tiny{$\pm 0.027$}  & $0.066$\tiny{$\pm 0.042$}  & $-0.102$\tiny{$\pm 0.110$} & $-0.169$\tiny{$\pm 0.083$} \\
            RF-ZIP        & $-0.265$\tiny{$\pm 0.014$} & $-0.222$\tiny{$\pm 0.018$} & $0.016$\tiny{$\pm 0.000$}  & $0.019$\tiny{$\pm 0.000$} \\
            Dos-DEL       & $-0.048$\tiny{$\pm 0.036$} & $-0.011$\tiny{$\pm 0.035$} & $-0.016$\tiny{$\pm 0.029$} & $-0.017$\tiny{$\pm 0.021$} \\
            DEL-QSVR      & $-0.228$\tiny{$\pm 0.021$} & $-0.171$\tiny{$\pm 0.033$} & $-0.004$\tiny{$\pm 0.178$} & $0.018$\tiny{$\pm 0.139$} \\
            DEL-Dock      & $-0.281$\tiny{$\pm 0.025$} & $-0.266$\tiny{$\pm 0.019$} & $-0.171$\tiny{$\pm 0.051$} & $-0.181$\tiny{$\pm 0.047$} \\
            DEL-Ranking   & $-0.289$\tiny{$\pm 0.012$} & $-0.245$\tiny{$\pm 0.009$} & $-0.177$\tiny{$\pm 0.034$} & $-0.193$\tiny{$\pm 0.027$} \\
            \bottomrule
        \end{tabular}
    }

    \vspace{0.2cm} 

    \resizebox{\linewidth}{!}{ 
        \begin{tabular}{@{}l c c c c@{}}
            \toprule
            & \multicolumn{2}{c}{\textbf{2hkf}} & \multicolumn{2}{c}{\textbf{5fl4}} \\
            \cmidrule(lr){2-3} \cmidrule(lr){4-5} 
            \textbf{Metric} & Sp & SubSp & Sp & SubSp \\
            \midrule
            Mol Weight    & -0.121 & -0.028 & -0.121 & -0.028 \\
            Benzene       & -0.174 & -0.134 & -0.174 & -0.134 \\
            \midrule
            Vina Docking  & $-0.114$\tiny{$\pm 0.009$} & $-0.055$\tiny{$\pm 0.007$} & $-0.114$\tiny{$\pm 0.007$} & $-0.055$\tiny{$\pm 0.006$} \\
            RF-Enrichment & $-0.016$\tiny{$\pm 0.021$} & $-0.014$\tiny{$\pm 0.030$} & $-0.064$\tiny{$\pm 0.126$} & $-0.144$\tiny{$\pm 0.024$} \\
            RF-ZIP        & $-0.053$\tiny{$\pm 0.017$} & $-0.066$\tiny{$\pm 0.034$} & $0.040$\tiny{$\pm 0.022$}  & $-0.011$\tiny{$\pm 0.042$} \\
            Dos-DEL       & $-0.003$\tiny{$\pm 0.030$} & $-0.048$\tiny{$\pm 0.034$} & $-0.115$\tiny{$\pm 0.065$} & $-0.036$\tiny{$\pm 0.010$} \\
            DEL-QSVR      & $0.070$\tiny{$\pm 0.134$}  & $-0.076$\tiny{$\pm 0.116$} & $-0.086$\tiny{$\pm 0.060$} & $-0.036$\tiny{$\pm 0.074$} \\
            DEL-Dock      & $-0.187$\tiny{$\pm 0.006$} & $-0.173$\tiny{$\pm 0.010$} & $-0.308$\tiny{$\pm 0.000$} & $-0.169$\tiny{$\pm 0.000$} \\
            DEL-Ranking   & $-0.190$\tiny{$\pm 0.005$} & $-0.155$\tiny{$\pm 0.009$} & $-0.323$\tiny{$\pm 0.015$} & $-0.175$\tiny{$\pm 0.000$} \\
            \bottomrule
        \end{tabular}
    }
\end{table}

\subsection{Top Case Selection}
\begin{figure}[t] 
    \centering
    \includegraphics[width=\linewidth]{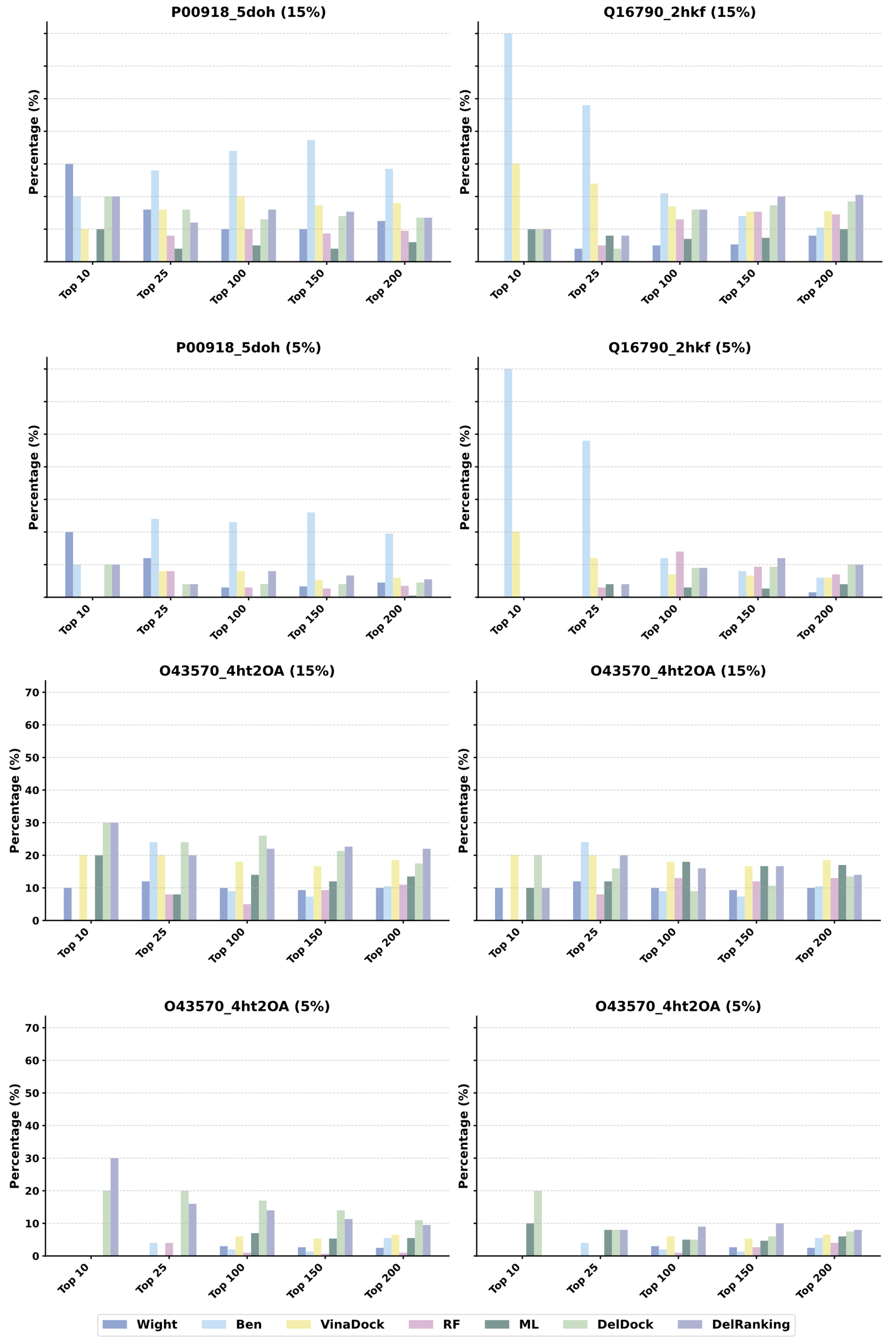}   
    
    \Description{A line chart (or bar chart) displaying the Top-N hit rate performance of various models.}
    
    \caption{Model performance evaluation using the Top-N hit rate on the CA-DEL dataset. This plot shows the percentage of high-affinity "hits" (defined as the top 5\% of binders) successfully identified within the top 200 predictions of each model's ranked list.}
    \label{fig:top}
\end{figure}

To assess the practical value of models in resource-constrained drug discovery campaigns, we introduce the Top-N Hit Rate analysis. Unlike global rank-order metrics (e.g., Spearman's $\rho$) that average performance across the dataset, this metric strictly evaluates "early enrichment"—specifically the percentage of confirmed high-affinity actives (top 5\% and 15\% $K_i$) identified within the top $N$ predicted compounds (Figure~\ref{fig:top}). The results reveal a critical dichotomy: while traditional 2D-based baselines exhibit sharp performance degradation in this regime, structure-aware deep learning models, notably DEL-Dock and DEL-Ranking, consistently dominate the Top-10 and Top-25 rankings. This indicates that 3D models effectively concentrate the most potent chemical matter at the very top of the list, thereby maximizing the return on investment for subsequent wet-lab synthesis.

This performance disparity provides profound insight into the nature of the learning task: while 2D topological features suffice for distinguishing broad classes of binders from non-binders, they lack the fidelity to discriminate between good binders and exceptional ones. The superior hit rates of 3D models suggest that the precise geometric constraints and stereochemical interactions governing high-affinity binding are successfully captured by the generated conformational ensembles. Ultimately, the clear performance hierarchy observed on CA-DEL—ascending from physicochemical heuristics to 2D-based ML, and peaking with 3D-integrated Deep Learning—validates the benchmark's premise. It demonstrates that overcoming the noise inherent in DEL screens and generalizing to OOD targets requires a transition from correlating statistical artifacts to learning fundamental, physically-grounded interaction features.


\subsection{Zero-shot Generalization}
To systematically evaluate out-of-distribution (OOD) generalization, our zero-shot experiments are constructed to assess model performance across two distinct levels of distribution shift: \textit{cross-conformation} and \textit{cross-target} transfer. Specifically, Table~\ref{tab:z1_5fl4_trained} presents models trained exclusively on the CAIX target using the \texttt{5fl4} crystal structure template. These models are then evaluated on an unseen conformation of CAIX (\texttt{2hkf}, representing cross-conformation transfer) as well as distinct, unseen homologous targets CAII (\texttt{3p3h}) and CAXII (\texttt{4kp5\_A} and \texttt{4kp5\_OA}, representing cross-target transfer). Conversely, Table~\ref{tab:z2_2hfk_trained} provides a symmetrical experimental setup where models trained on CAIX using the \texttt{2hkf} template are evaluated for cross-conformation transfer to \texttt{5fl4} and cross-target transfer to CAII and CAXII.

The results reveal clear boundaries in the capabilities of current structure-based models. While the models demonstrate competence within their in-distribution structural templates, their performance degrades significantly when forced to generalize across unseen protein states or homologous targets. A notable example is the cross-target transfer from \texttt{5fl4} (trained on CAIX) to \texttt{4kp5\_OA} (tested on CAXII), where the baseline model yields a positive Spearman correlation of $0.065 \pm 0.021$. In our Sim-to-Real evaluation framework—where a strong negative SubSp ($\rho \rightarrow -1$) signifies accurate $K_i$ affinity ranking—this positive value indicates a systematic inversion of the predictive ranking. Effectively, the model ranks experimentally validated binders lower than non-binders in this specific OOD scenario.

This outcome highlights a key challenge validated by the CA-DEL benchmark: neither homologous target similarity (e.g., CAIX to CAXII) nor identical target conservation (e.g., cross-conformation transfer within CAIX) is sufficient to guarantee generalization when accompanied by substantial distributional shifts in ligand space (from DEL enrichment to ChEMBL $K_i$). The observation that models struggle to maintain predictive power even when transferring between distinct crystal structures of the \textit{same} target isoform (e.g., \texttt{5fl4} $\leftrightarrow$ \texttt{2hkf}) suggests that current 3D models are highly sensitive to conformational variations. Rather than fully capturing the underlying physics-based binding mechanisms, they tend to overfit to the specific rigid conformational state of the training template and rely on library-specific statistical features inherent to the source DEL data. By exposing these dual challenges—conformational sensitivity and cross-target shortcut learning—CA-DEL serves as a rigorous testbed to guide the development of more robust and generalizable geometric deep learning models.

\begin{table}[t]
    \centering
    \small
    
    \caption{Cross-conformation and cross-target zero-shot generalization performance. Models were trained exclusively on the \textbf{Carbonic Anhydrase IX (CAIX)} target using crystal structure \textbf{5fl4}.
    Top: Zero-shot transfer to the \textbf{CAII} target (3p3h) and \textbf{CAIX} (2hkf). 
    Bottom: Zero-shot transfer to \textbf{CAXII} targets (4kp5\_A and 4kp5\_OA).}
    \label{tab:z1_5fl4_trained}

    \resizebox{\linewidth}{!}{
        \begin{tabular}{@{}lcccc@{}}
        \toprule
        & \multicolumn{2}{c}{\textbf{3p3h}} & \multicolumn{2}{c}{\textbf{2hkf}} \\
        \cmidrule(lr){2-3} \cmidrule(lr){4-5}
        & Sp & SubSp & Sp & SubSp \\
        \midrule
        DEL-Dock    & -0.272$\pm$\tiny{0.013} & -0.118$\pm$\tiny{0.005} & -0.108$\pm$\tiny{0.011} & -0.110$\pm$\tiny{0.019} \\
        DEL-Ranking & -0.310$\pm$\tiny{0.005} & -0.120$\pm$\tiny{0.011} & -0.218$\pm$\tiny{0.017} & -0.177$\pm$\tiny{0.009} \\
        \bottomrule
        \end{tabular}
    }

    \vspace{0.2cm} 

    \resizebox{\linewidth}{!}{
        \begin{tabular}{@{}lcccc@{}}
        \toprule
        & \multicolumn{2}{c}{\textbf{4kp5\_A}} & \multicolumn{2}{c}{\textbf{4kp5\_OA}} \\
        \cmidrule(lr){2-3} \cmidrule(lr){4-5}
        & Sp & SubSp & Sp & SubSp \\
        \midrule
        DEL-Dock    & -0.211$\pm$\tiny{0.007} & -0.118$\pm$\tiny{0.010} & 0.065$\pm$\tiny{0.021} & -0.125$\pm$\tiny{0.034} \\
        DEL-Ranking & -0.228$\pm$\tiny{0.010} & -0.127$\pm$\tiny{0.018} & -0.300$\pm$\tiny{0.026} & -0.129$\pm$\tiny{0.021} \\
        \bottomrule
        \end{tabular}
    }
\end{table}
\begin{table}[t]
    \centering
    \small 
    
    \caption{Cross-conformation and cross-target zero-shot generalization performance. Models were trained exclusively on the \textbf{Carbonic Anhydrase IX (CAIX)} target using crystal structure \textbf{2hkf}.  
    Top: Zero-shot transfer to the \textbf{CAII} target (3p3h) and \textbf{CAIX} (5fl4). 
    Bottom: Zero-shot transfer to \textbf{CAXII} targets (4kp5\_A and 4kp5\_OA).}
    \label{tab:z2_2hfk_trained}

    \resizebox{\linewidth}{!}{
        \begin{tabular}{@{}lcccc@{}}
        \toprule
        & \multicolumn{2}{c}{\textbf{3p3h}} & \multicolumn{2}{c}{\textbf{5fl4}} \\
        \cmidrule(lr){2-3} \cmidrule(lr){4-5}
        & Sp & SubSp & Sp & SubSp \\
        \midrule
        DEL-Dock    & -0.185$\pm$\tiny{0.016} & -0.166$\pm$\tiny{0.008} & -0.185$\pm$\tiny{0.012} & -0.162$\pm$\tiny{0.021} \\
        DEL-Ranking & -0.224$\pm$\tiny{0.011} & -0.209$\pm$\tiny{0.018} & -0.150$\pm$\tiny{0.015} & -0.095$\pm$\tiny{0.010} \\
        \bottomrule
        \end{tabular}
    }

    \vspace{0.2cm} 

    \resizebox{\linewidth}{!}{
        \begin{tabular}{@{}lcccc@{}}
        \toprule
        & \multicolumn{2}{c}{\textbf{4kp5\_A}} & \multicolumn{2}{c}{\textbf{4kp5\_OA}} \\
        \cmidrule(lr){2-3} \cmidrule(lr){4-5}
        & Sp & SubSp & Sp & SubSp \\
        \midrule
        DEL-Dock    & -0.118$\pm$\tiny{0.009} & -0.062$\pm$\tiny{0.014} & 0.048$\pm$\tiny{0.011} & 0.043$\pm$\tiny{0.007} \\
        DEL-Ranking & -0.174$\pm$\tiny{0.023} & -0.124$\pm$\tiny{0.006} & -0.091$\pm$\tiny{0.019} & -0.044$\pm$\tiny{0.013} \\
        \bottomrule
        \end{tabular}
    }
\end{table}

\subsection{Baseline Models and Experimental Setup}
To provide a solid performance reference for future research, we established a suite of baseline models spanning various levels of complexity and data modalities. The selection of these models is intended to systematically probe the capabilities and limitations of different methodologies in addressing the unique challenges presented by the CA-DEL dataset. Our library of baseline models extends from simple physicochemical prior-based methods to advanced multi-modal deep learning models, and specifically includes:

\begin{itemize}
    \item\textbf{Training DEL for Ranking Targets:} This task focuses on training the model to distinguish active compounds from decoys, effectively solving the hit-finding problem through classification and ranking.

    \item\textbf{Tuning DEL with $K_i/K_d$ for Targets Affinity Prediction:} Here, the model is fine-tuned for a quantitative regression task to predict the precise binding affinity of molecules to their biological targets.

    \item\textbf{Generalizing DEL for New Targets:} This task evaluates the model's ability to transfer its learned knowledge and make accurate predictions for novel protein targets not seen during training.

\end{itemize}
All experiments were completed on a computing cluster equipped with NVIDIA A100 GPUs. We specifically note that the computational cost of the 3D models is significantly higher than that of the 2D models due to the need to process multiple 3D conformations for each molecule, a fact that highlights the importance of releasing pre-computed features and standardized benchmarks. Finally, to promote full reproducibility and further research by the community, all associated code and data will be made publicly available through a GitHub repository and Zenodo.

\section{Conclusion}
In this work, we introduced CA-DEL, a comprehensive multi-target, multi-modal benchmark designed to bridge the critical gap between noisy high-throughput DEL screening signals and validated biophysical reality. By integrating large-scale enrichment data with high-fidelity $K_i$ measurements across homologous Carbonic Anhydrase isoforms, we have established a rigorous evaluation paradigm that mirrors the actual challenges of hit-to-lead optimization. Our extensive experimental analysis demonstrates that models leveraging 3D conformational ensembles significantly outperform traditional 2D baselines in both correlation with true binding affinities and practical Top-N hit retrieval, validating the hypothesis that explicit geometric modeling is essential for effective denoising of proxy sequencing signals. However, the observed limitations in zero-shot generalization and fine-grained isoform selectivity highlight that current state-of-the-art approaches still rely heavily on dataset-specific artifacts, underscoring the urgent need for future research to focus on physics-informed architectures, uncertainty quantification, and explainable AI to fully realize the potential of rational drug design.

\subsection{GenAI Disclosure}
Large Language Models (LLMs) were used to aid in the writing and polishing of the manuscript. Specifically, we used an LLM to assist in refining the language, improving readability, and ensuring clarity in various sections of the paper. The model helped with tasks such as sentence rephrasing, grammar checking, and enhancing the overall flow of the text.

It is important to note that the LLM was not involved in the ideation, research methodology, or experimental design. All research concepts, ideas, and analyses were developed and conducted by the authors. The contributions of the LLM were solely focused on improving the linguistic quality of the paper, with no involvement in the scientific content or data analysis.

The authors take full responsibility for the content of the manuscript, including any text generated or polished by the LLM. We have ensured that the LLM-generated text adheres to ethical guidelines and does not contribute to plagiarism or scientific misconduct.

\nobalance 
\clearpage        
\bibliographystyle{ACM-Reference-Format}
\bibliography{iclr2026/iclr2026_conference}

\clearpage        
\onecolumn        
\appendix         
\renewcommand{\thefigure}{A\arabic{figure}}
\renewcommand{\thetable}{A\arabic{table}}
\setcounter{figure}{0}
\setcounter{table}{0}

\section{Appendix}
\subsection{Appendix figure and Table}
\begin{figure}[H]
    \centering
    \includegraphics[width=0.8\linewidth]{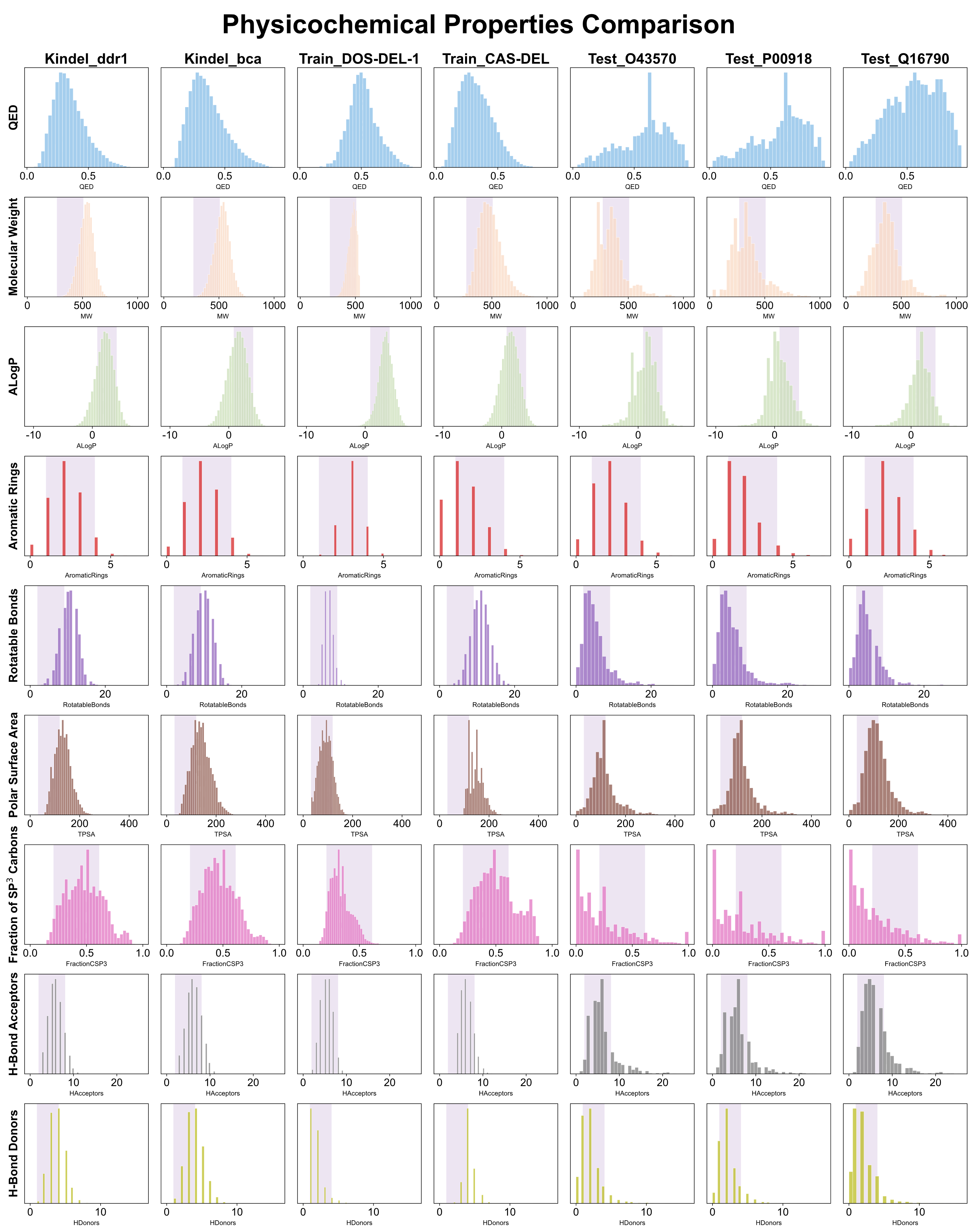}
    \caption{
        Comparison of key physicochemical properties (QED, LogP, Molecular Weight and others) across distinct ligand datasets, which includes source datasets (DOS-DEL-1 and CAS-DEL), benchmark dataset KinDEL~\citep{chen2024kindel}, and our curated test sets.The light blue areas mark the 10th and 90th percentiles computed for all the FDA approved oral new chemical entities, as reported by ~\citep{shultz2018two}. QED: quantitative estimate of druglikeness~\citep{bickerton2012quantifying}.
    }
    \label{fig:physchem_props2}
\end{figure}

\begin{figure}[H]
    \centering
    \includegraphics[width=0.99\linewidth]{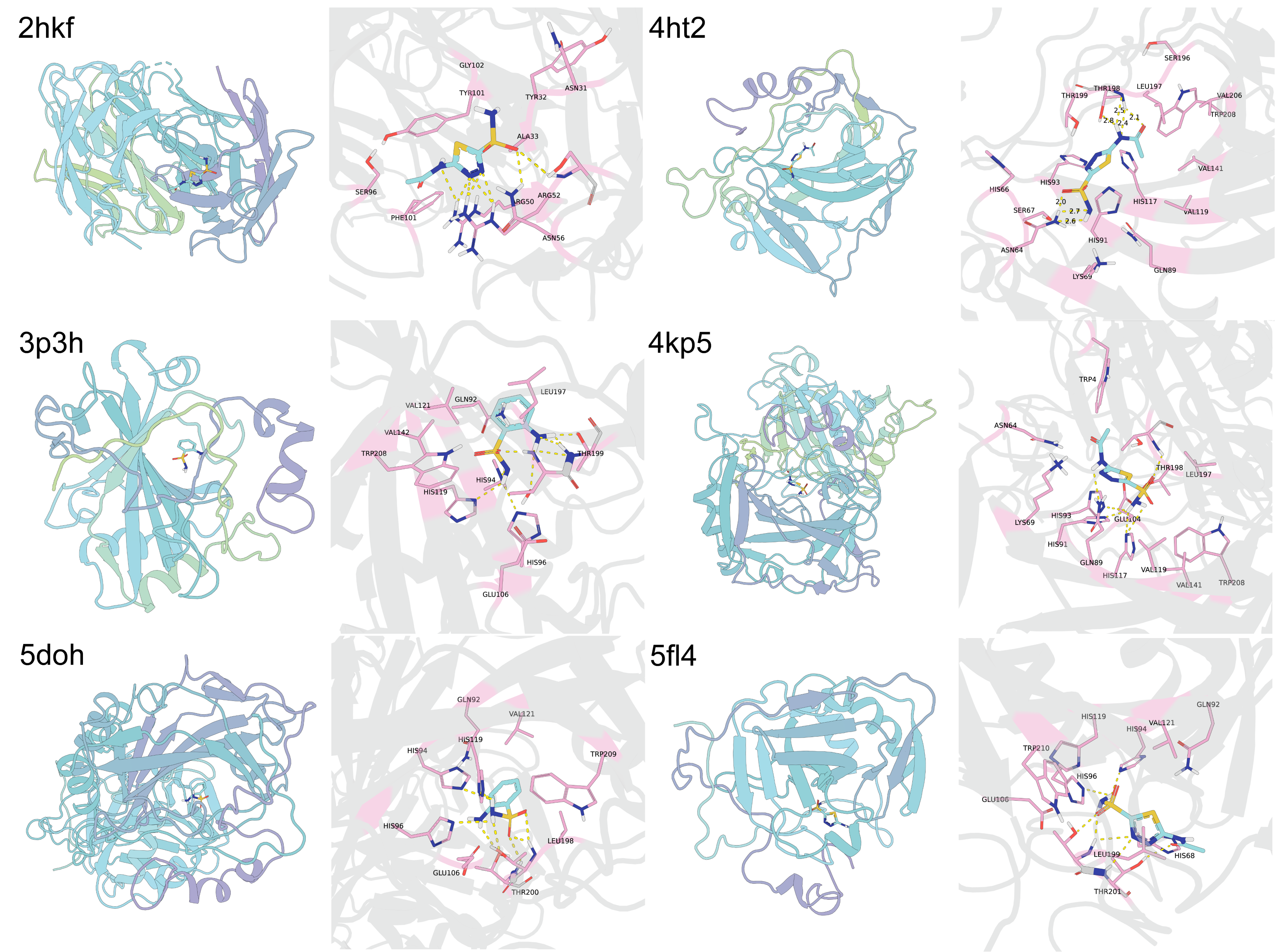}
    \caption{Validation of the docking protocol via re-docking. \textbf{(Left)} Overall view of the computationally generated binding pose (ligand shown in cyan sticks) within the protein's active site. \textbf{(Right)} A zoomed-in view detailing the critical interactions. The computationally generated pose (cyan) accurately reproduces the experimental binding mode(such as zinc chelation and hydrogen bonds), forming key interactions with the active site residues (pink sticks).}
    \label{fig:re-docking}
\end{figure}

\begin{longtable}{@{} >{\ttfamily}p{0.22\textwidth} p{0.2\textwidth} l >{\raggedright\arraybackslash}p{0.32\textwidth}@{}}
\caption{Description of data fields in the CAS-DEL dataset files. This table can span multiple pages.} \\
\label{tab:appendix_data_fields_cas} \\
\toprule
\textbf{Field Name} & \textbf{Data Type} & \textbf{Source} & \textbf{Description} \\
\midrule
\endfirsthead
\caption{(continued) Description of data fields in the CAS-DEL dataset files.} \\
\toprule
\textbf{Field Name} & \textbf{Data Type} & \textbf{Source} & \textbf{Description} \\
\midrule
\endhead
\bottomrule
\endfoot
smiles & String & CAS-DEL & Simplified Molecular-Input Line-Entry System (SMILES) string representing the final molecule. \\
\addlinespace
CodeA, CodeB, CodeC & String / Categorical & CAS-DEL & Unique identifiers for the DNA barcodes or chemical building blocks used in each of the three synthesis rounds. \\
\addlinespace
Pre & Integer & CAS-DEL & Pre-selection read counts. \\
\addlinespace
P & Integer & CAS-DEL & Raw read counts of DNA tags obtained after the panning step (incubation with the target protein). \\
\addlinespace
A & Integer & CAS-DEL & Raw read counts of DNA tags obtained after the amplification step (PCR). \\
\addlinespace
OA & Integer & CAS-DEL & Raw read counts of DNA tags obtained after the off-target / counter-selection step. \\
\addlinespace
Post & Integer & CAS-DEL & Final enriched DNA tag raw read counts obtained after the complete selection process. \\
\addlinespace
label & Integer / Boolean & CAS-DEL & Binary classification label (e.g., 1 for active, 0 for inactive) assigned based on a preset enrichment threshold or validation experiment results. Serves as the target variable for predictive models. \\
\addlinespace
row\_count & Integer & CAS-DEL & Row index. \\
\addlinespace
Exp-B01 & Integer & CAS-DEL & Blank control counts: Raw NGS reads from a control selection performed with blank magnetic beads (no immobilized target protein). \\
\addlinespace
Exp-P01, P02, P03 & Integer & CAS-DEL & Post-selection counts (purified protein): Raw NGS reads from three biological replicates of selections against purified, immobilized Carbonic Anhydrase II (CAII). \\
\addlinespace
Exp-A01, A02, A03 & Integer & CAS-DEL & Post-selection counts (endogenous cellular target): Raw NGS reads from three biological replicates of selections against A549 cells expressing endogenous levels of membrane-bound Carbonic Anhydrase XII (CAXII). \\
\addlinespace
Exp-OA01, OA02, OA03 & Integer & CAS-DEL & Post-selection counts (overexpressed cellular target): Raw NGS reads from three biological replicates of selections against hypoxic A549 cells overexpressing Carbonic Anhydrase XII (CAXII). \\
\addlinespace
number & Integer & CAS-DEL & Compound serial number. \\
\addlinespace
SMILES\_BB1, BB2, BB3 & String & CAS-DEL & SMILES strings for the individual Building Blocks (BBs) used to construct the final molecule. \\
\addlinespace
BB1D, BB2D, BB3D & String / Categorical & CAS-DEL & Descriptor or ID for the corresponding building blocks. \\
\bottomrule
\end{longtable}

\begin{longtable}{@{} >{\ttfamily}p{0.22\textwidth} p{0.2\textwidth} l >{\raggedright\arraybackslash}p{0.32\textwidth}@{}}
\caption{Description of data fields in the DOS-DEL-1 dataset. This table can span multiple pages.} \\
\label{tab:appendix_data_fields_dos} \\
\toprule
\textbf{Field Name} & \textbf{Data Type} & \textbf{Source} & \textbf{Description} \\
\midrule
\endfirsthead
\caption{(continued) Description of data fields in the DOS-DEL-1 dataset.} \\
\toprule
\textbf{Field Name} & \textbf{Data Type} & \textbf{Source} & \textbf{Description} \\
\midrule
\endhead
\bottomrule
\endfoot
(Unnamed: 0) & Integer & DOS-DEL-1 & Row index. \\
\addlinespace
cpd\_id & String / Integer & DOS-DEL-1 & Unique identifier for each compound. \\
\addlinespace
scaffold, BB1, BB2 & String / Integer & DOS-DEL-1 & Identifiers for the chemical scaffold and the two variable building blocks (BB1, BB2) used in the DOS-DEL library synthesis. \\
\addlinespace
ap1\_baseline & Integer & DOS-DEL-1 & Baseline read counts; initial sequencing counts for a CAIX-specific cell line. \\
\addlinespace
hrp\_beads\_r1...r4 & Integer & DOS-DEL-1 & Counter-selection / off-target counts: Raw NGS reads from replicate selections against an unrelated protein (Horseradish Peroxidase, HRP). \\
\addlinespace
ca9\_beads\_r1...r2 & Integer & DOS-DEL-1 & On-target selection counts: Raw NGS reads from replicate selections against the immobilized target protein, Carbonic Anhydrase IX (CAIX). \\
\addlinespace
hrp\_exp\_r1...r2 & Integer & DOS-DEL-1 & Raw NGS reads from replicate experimental output pools in the HRP counter-selection experiment. \\
\addlinespace
ca9\_exp\_r1...r4 & Integer & DOS-DEL-1 & Raw NGS reads from replicate experimental output pools in the CAIX on-target selection experiment. \\
\addlinespace
hrp\_B, A, Bp, Ap & Float & DOS-DEL-1 & Calculated metrics for HRP counter-selection. Represents: Baseline (B), After-selection (A), Baseline percentage (Bp), and After-selection percentage (Ap). \\
\addlinespace
ca9\_B, A, Bp, Ap & Float & DOS-DEL-1 & Calculated metrics for CAIX on-target selection, with meanings as described above. \\
\addlinespace
ca9\_Fn, hrp\_Fn & Float & DOS-DEL-1 & Normalized fold-change (Fn): Statistically corrected enrichment scores calculated for the on-target (CAIX) and counter-selection (HRP) screens, respectively. \\
\addlinespace
scaffold\_smiles, ... & String & DOS-DEL-1 & SMILES strings for the chemical scaffold, building blocks, and the final combined molecule. \\
\addlinespace
collection, type & String / Cat. & DOS-DEL-1 & Collection or type metadata for the chemical library. \\
\addlinespace
ecfp6 & String / Bit Vector & DOS-DEL-1 & Extended-Connectivity Fingerprint (diameter 6). A hashed numerical representation of molecular structure. \\
\addlinespace
cycle0, cycle1, cycle2 & Integer & DOS-DEL-1 & Post-selection round counts: Raw NGS reads after zero, one, or two cycles of selection-amplification. \\
\addlinespace
library\_id & String / Integer & DOS-DEL-1 & Identifier for the DNA-Encoded Library used. \\
\addlinespace
cycle01, cycle02, cycle12 & Float & DOS-DEL-1 & Enrichment ratios between different selection rounds (e.g., cycle1/cycle0), providing a direct measure of a compound's enrichment efficiency. \\
\bottomrule
\end{longtable}

\begin{longtable}{@{} >{\ttfamily}p{0.22\textwidth} p{0.2\textwidth} l >{\raggedright\arraybackslash}p{0.32\textwidth}@{}}
\caption{Description of data fields for the ChEMBL validation and test sets.} \\
\label{tab:appendix_data_fields_chembl} \\
\toprule
\textbf{Field Name} & \textbf{Data Type} & \textbf{Source} & \textbf{Description} \\
\midrule
\endfirsthead
\caption{(continued) Description of data fields for the ChEMBL validation and test sets.} \\
\toprule
\textbf{Field Name} & \textbf{Data Type} & \textbf{Source} & \textbf{Description} \\
\midrule
\endhead
\bottomrule
\endfoot
num & Integer & ChEMBL & Row index. \\
\addlinespace
smiles & String & ChEMBL & Simplified Molecular-Input Line-Entry System (SMILES) string. \\
\addlinespace
affinity & Float / String & ChEMBL & The specific type of affinity measurement reported (e.g., Ki, Kd, IC50). \\
\addlinespace
Ki (nM) & Float & ChEMBL & Inhibition constant, reported in nanomolar (nM) units. \\
\addlinespace
Kd (nM) & Float & ChEMBL & Dissociation constant, reported in nanomolar (nM) units. \\
\addlinespace
IC50 (nM) & Float & ChEMBL & Half-maximal inhibitory concentration, reported in nanomolar (nM) units. \\
\bottomrule
\end{longtable}

\subsection{Detailed Protocols and Experimental Setup}
To facilitate fair, reproducible, and rigorous evaluation on our CA-DEL dataset, we have unified our benchmark suite into three core tasks. These tasks directly align with our main experimental results and probe model capabilities across in-distribution ranking, Sim-to-Real generalization, and out-of-distribution (OOD) cross-target transfer.
\begin{itemize}
    \item \textbf{In-Distribution Ranking (DEL-to-DEL).} This task evaluates a model's ability to learn from noisy, high-throughput DEL signals. The objective is to train a model that takes molecular information (2D and/or 3D representations) as input and outputs a continuous ranking score that monotonically correlates with the experimental sequencing read counts. Performance is measured by the Spearman's rank correlation coefficient (Spearman's $\rho$) on the held-out DEL test sets. This assesses the model's baseline ability to fit the experimental proxy signal.

    \item \textbf{Sim-to-Real Generalization (DEL-to-ChEMBL).} This is the defining OOD challenge of CA-DEL, bridging the gap between noisy screening data (Sim) and precise biophysical measurements (Real). Models trained strictly on DEL enrichment signals are evaluated on the rigorous ChEMBL test set containing ground-truth experimental binding affinities. Performance is evaluated using two practical metrics:
    \begin{itemize}
        \item \textit{SubSp (Sub-Spearman's $\rho$)}: The correlation between the model's predicted scores and the true binding affinities (e.g., $K_i$) on the ChEMBL set. \textbf{Note:} Because a lower $K_i$ indicates tighter binding affinity, a stronger \textit{negative} SubSp indicates better predictive performance.
        \item \textit{Top-N Hit Rate}: Measuring the proportion of experimentally validated true binders recovered within the model's top $N$ ranked predictions. This metric directly simulates real-world hit-triage utility.
    \end{itemize}
    \item \textbf{Cross-Conformation and Cross-Target Zero-Shot Generalization.} This task aims to evaluate a model's ability to learn transferable chemical and structural knowledge across varying structural states and homologous protein isoforms. To strictly align with our main experimental settings (e.g., Tables 4 and 5), the zero-shot protocol evaluates models on two levels of distribution shift. Specifically, a model is trained exclusively on data derived from a single crystal structure template of a specific target (e.g., the CAIX target using PDB ID 5fl4 or 2hkf). It is then directly applied, without fine-tuning, to make predictions on datasets generated using either (1) an unseen conformation of the \textit{same} target (cross-conformation, e.g., transferring from CAIX 5fl4 to CAIX 2hkf) or (2) distinct, unseen homologous targets (cross-target, e.g., transferring from CAIX to CAII 3p3h or CAXII 4kp5). Performance is measured by the Spearman's rank correlation coefficient ($\rho$) between the predicted scores and the experimental read counts on the unseen sets. By explicitly mapping specific crystal templates to their respective target isoforms, this protocol rigorously exposes the fundamental limits of zero-shot transferability against both conformational flexibility and target homology within the CA family.
    \item \textbf{Advanced Challenge: Explicit Selectivity Prediction}
\end{itemize}

To ensure the fairness of all comparisons and the reproducibility of results, we provide standard, fixed data splits for CA-DEL (in \texttt{.npz} format). For each sub-dataset, we provide splits for training (80\%), validation (10\%), and testing (10\%). We strongly encourage researchers to adhere to these prescribed splits. All model hyperparameters should be tuned on the validation set, with final performance reported on the independent test set. All code for data processing, task evaluation, and baseline model implementation will be publicly released to support this protocol.

\subsection{model size and parameterization details}
In this section, we present the relevant hyperparameters.  All hyperparameters were tuned to achieve optimal performance on their respective datasets.
\begin{table}[H]
    \centering
    \caption{DEL-Docking Hyperparameters (Model Structure Parameters)}
    \label{tab:del-ranking-hyper-2}
    \begin{tabular}{lcccc}
        \toprule
        Dataset & learning\_rate & lrd\_gamma & n\_layers & dropout \\
        \midrule
        2hkf   & 3.00E-04 & 0.1 & 2 & 0.5 \\
        3p3h   & 8.00E-06 & 0.1 & 2 & 0.5 \\
        4ht2A  & 2.00E-05 & 0.1 & 2 & 0.5 \\
        4ht2OA & 2.00E-05 & 0.1 & 2 & 0.5 \\
        4kp5A  & 8.00E-05 & 0.1 & 2 & 0.5 \\
        4kp5OA & 1.00E-04 & 0.1 & 2 & 0.5 \\
        5doh   & 5.00E-06 & 0.1 & 2 & 0.5 \\
        5fl4   & 1.00E-04 & 0.1 & 2 & 0.5 \\
        \bottomrule
    \end{tabular}
\end{table}
\begin{table}[H]
    \centering
    \caption{DEL-Ranking Hyperparameters (Optimization Parameters)}
    \label{tab:del-ranking-hyper-1}
    \begin{tabular}{lcccc}
        \toprule
        Dataset & lr & listmle\_weight & $\lambda_w$ & Temperature \\
        \midrule
        2hkf   & 8.00E-05 & 1.00E-09 & 1.00E+09 & 0.1 \\
        3p3h   & 3.00E-05 & 1.00E-09 & 1.00E+09 & 0.5 \\
        4ht2A  & 3.00E-05 & 1.00E-09 & 1.00E+09 & 0.8 \\
        4ht2OA & 3.00E-05 & 1.00E-09 & 1.00E+09 & 0.8 \\
        4kp5A  & 1.00E-03 & 1.00E-09 & 1.00E+09 & 0.3 \\
        4kp5OA & 1.00E-03 & 1.00E-10 & 1.00E+10 & 0.2 \\
        5doh   & 1.00E-04 & 1.00E-10 & 1.00E+10 & 0.8 \\
        5fl4   & 1.00E-04 & 1.00E-08 & 1.00E+08 & 0.9 \\
        \bottomrule
    \end{tabular}
\end{table}

\subsection{Data Availability}
The dataset has been uploaded to Zenodo under the CC BY 4.0 license (\url{https://zenodo.org/records/17656024}). This repository ensures the data is Findable (via DOI), Accessible (open download), Interoperable (standard.sdf and.pdb formats), and Reusable (clear licensing), strictly adhering to FAIR principles.

\end{document}